\documentclass[acp, manuscript]{copernicus}

\usepackage{latexsym,euscript,textcomp}
\usepackage{color,graphics,epsf,dcolumn}
\usepackage{epstopdf,ulem}

\begin{document}
\nolinenumbers

\title{Comments on ``Is condensation-induced atmospheric dynamics
a new theory of the origin of the winds?" by \citet{jaramillo18}}


\Author[1,2]{Anastassia M.}{Makarieva}
\Author[1,2]{Victor G.}{Gorshkov}
\Author[3]{Antonio Donato}{Nobre}
\Author[1]{Andrei V.}{Nefiodov}
\Author[4]{Douglas}{Sheil}
\Author[5]{Paulo}{Nobre}
\Author[2]{Bai-Lian}{Li}

\affil[1]{Theoretical Physics Division, Petersburg Nuclear Physics Institute, Gatchina  188300, St.~Petersburg, Russia}
\affil[2]{USDA-China MOST Joint Research Center for AgroEcology and Sustainability, University of California, Riverside 92521-0124, USA}
\affil[3]{Centro de Ci\^{e}ncia do Sistema Terrestre INPE, S\~{a}o Jos\'{e} dos Campos, S\~{a}o Paulo  12227-010, Brazil}
\affil[4]{Faculty of Environmental Sciences and Natural Resource Management, Norwegian University of Life Sciences, \AA s, Norway}
\affil[5]{Center for Weather Forecast and Climate Studies INPE, S\~{a}o Jos\'{e} dos Campos, S\~{a}o Paulo 12227-010, Brazil}


\runningtitle{Comments on ``Is condensation-induced atmospheric dynamics
a new theory of the origin of the winds?" by \citet{jaramillo18} }

\runningauthor{Makarieva et al.}

\correspondence{Anastassia Makarieva (ammakarieva@gmail.com)}

\received{}
\pubdiscuss{} 
\revised{}
\accepted{}
\published{}


\firstpage{1}

\maketitle


\introduction  

\label{intr}

\citet{jaramillo18} criticized our theory of condensation-induced atmospheric dynamics (CIAD). We value any such interest but, as we show below, \citet{jaramillo18}'s main statement, that CIAD modifies the equation of vertical motion such that it violates Newton's third law, is unsupported. Contrary to their claims, CIAD does not make any {\it ``modification to the vertical momentum budget"} (correct or incorrect) nor
to any fundamental equations of hydrodynamics.  (Despite claiming their assessment to be {\it ``rigorous"}, \citet{jaramillo18} don't locate the alleged equation in our publications.) Rather, as we summarize below, CIAD constrains the {\it power of atmospheric circulation} in a manner that is consistent with observations.

More specifically, we find that \citet{jaramillo18}'s analysis of the equation of vertical motion
is invalid: it confuses the internal and external forces acting on a unit volume of air.
The equations of motion for moist air are complicated \citep{trenberth18}, so confusions do occur.
For example, the inconsistency between two conflicting equations of motions published within one year
in same meteorological journal had persisted unresolved for over 15 years \citep{oo01,ba02}.
That controversy related to misinterpretations of Newton's third law \citep[see][their Fig.~1]{jgra17},
which \citet{jaramillo18} perpetuate.

\citet{jaramillo18} correctly note that two different expressions for the evaporative force $f_e$, a key element of CIAD,
occur in our publications. We use this opportunity to clarify these expressions.
Finally, we clarify that there is no disagreement between CIAD and consideration of the atmosphere as a heat engine.
Contrary to the claims of \citet{jaramillo18}, these approaches address different problems and are complementary.

\section{The equation of vertical motion}

\citet{jaramillo18}'s statement can be summarized as follows. They write the equation of vertical motion
as
\begin{align} \label{vm}
F_z  &= \left(-\frac{\partial p_d}{\partial z} - g\rho_d + F_{vd}\right) + \left(-\frac{\partial p_v}{\partial z} - g\rho_v + F_{dv}\right) \nonumber \\
&=-\frac{\partial p}{\partial z} - g\rho + F_{vd}+ F_{dv} = \rho a_z,
\end{align}
where $a_z$ is the vertical acceleration of air, $g$ is the acceleration of gravity, $p_d$, $p_v$, $p = p_d+p_v$ and $\rho_d$,
$\rho_v$, $\rho = \rho_d+\rho_v$ denote, respectively, pressure and density of dry air, water vapor and moist air as a whole.
The terms grouped in parentheses  are interpreted by \citet{jaramillo18} as {\it ``the forces on each component"}
-- dry air and water vapor. Accordingly, forces $F_{vd}$ and $F_{dv}$ are defined
as {\it ``respectively the force of the vapor on the dry air and the force of  the dry air on the vapor, as mediated by molecular
collisions between the two components"}. According to \citet{jaramillo18}, these {\it ``internal forces"} must cancel due
to Newton's third law, $F_{dv} = -F_{vd}$.

\citet{jaramillo18} further state that {\it ``if  the air parcel is not undergoing vertical acceleration, then
$F_{vd} = f_e$, as defined by (6)"}. In their notations this means that
\begin{equation}\label{fvd}
F_{vd} = f_e \equiv -\frac{\partial p_v}{\partial z} - \frac{p_v}{h_v} = -\frac{\partial p_v}{\partial z} - \rho_v g,
\end{equation}
since $h_v \equiv RT /M_v g$, $p_v = N_vRT$ (ideal gas law), and $\rho_v = M_vN_v$, where
$R$ is the ideal gas constant, $T$ is temperature, $N_v$ and $M_v$ are molar density and molar mass of water vapor.
From this statement, \citet{jaramillo18}  proceed directly to their conclusion that {\it ``the flaw [in CIAD] is now clear"}:
it {\it ``includes  $F_{vd}$ in the vertical motion equation while omitting $F_{dv}$"}, which represents
{\it ``a clear violation of Newton's third law"}.

However, $F_{dv}$ and $F_{vd}$ cancel and thus cannot be retrieved from Eq.~(\ref{vm}).
Despite claiming their approach to be {\it ``rigorous"}, \citet{jaramillo18} themselves do not explain
how their central statement -- Eq.~(\ref{fvd}) -- was obtained. We speculate that they might have separated the
equation of motion (\ref{vm}) into two {\it ``component"} equations, for water vapor and dry air,
\begin{align}
\label{ev}
\rho_v a_{zv} &= -\frac{\partial p_v}{\partial z} - \rho_v g + F_{dv},  \\
\rho_d a_{zd} &= -\frac{\partial p_d}{\partial z} - \rho_d g + F_{vd},   \label{ed}
\end{align}
where $a_{zv}$ and $a_{zd}$ are vertical accelerations of water vapor and dry air.
Then {\it ``if   the air parcel is not undergoing vertical acceleration"}, $a_{zv} = a_{zd} = 0$ and
Eq.~(\ref{fvd}) follows from Eq.~(\ref{ev}) and the assumed $F_{dv} = -F_{vd}$.

\begin{figure}[tb]
\vspace{-0.5 cm}
\begin{minipage}[p]{\textwidth}
\centering\includegraphics[width=0.8\textwidth,angle=0,clip]{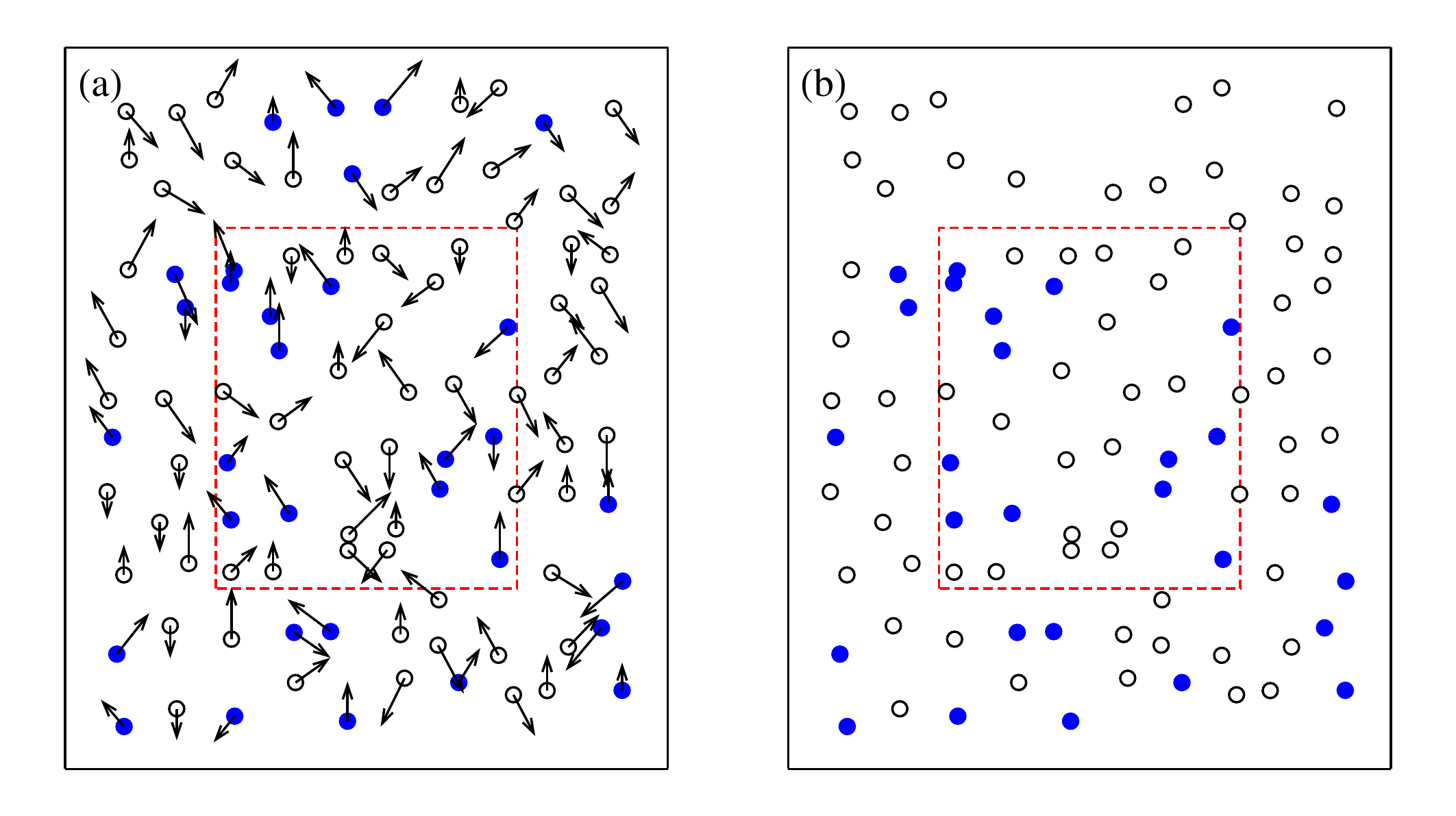}
\end{minipage}
\caption{Momentum exchange between gas molecules (open circles -- dry air, filled circles -- water vapor, dashed frame
denotes the considered unit volume).
(a) The cartoon is a reminder that all types of molecules collide with each other (arrows show the chaotic
velocities of molecular motion) and there is no rule that dry air (water vapor) outside the volume only collides with dry air (water vapor) within the volume
as implied by Eqs.~(\ref{ev}) and (\ref{ed}); (b)
The gradient of water vapor is perturbed from the initial
equilibrium state (a) by an instantaneous removal of water vapor from the upper forth of the vessel;
the gradient of dry air is not perturbed; within the unit volume nothing changes either -- in particular, interactions between the molecules remain the same; for clarity, molecular velocities
are not shown. In this case, according to Eqs.~(\ref{ev}) and (\ref{ed}), only water vapor will accelerate upward to fill the void, while the dry air as a whole will remain motionless. This absurd conclusion results from the incorrect identification of external forces in Eqs.~(\ref{ev}) and (\ref{ed}).}
\label{fig1}
\end{figure}

The problem with this assumed derivation is that Eqs.~(\ref{ev}) and (\ref{ed}) are incorrect. It is possible to write separate
equations of motion for such components of moist air as gas and condensate including their mutual interaction governed by
Newton's third law, see, e.g., Eq.~(23) and Fig.~1 of \citet{jgra17}.
This exercise requires, however, a correct identification of the external forces acting on the two components.
In the case of Eqs.~(\ref{ev}) and (\ref{ed}), while it is true that gravity acts separately on dry
air and water vapor and is, respectively,  $\rho_d g$ and $\rho_v g$,
it is an error to assume that $\partial p_d/\partial z$, the partial pressure gradient of dry air, acts exclusively on dry air,
while the partial pressure gradient of water vapor, $\partial p_v/\partial z$, acts exclusively on water vapor (Fig.~\ref{fig1}).

Borrowing the words of \citet{jaramillo18}, {\it ``as mediated by molecular collisions between the two components"},
these forces are not separable. The total pressure gradient $\partial p/\partial z$ is an external force acting on a unit volume of air.
Molecules of all gases adjacent to the volume collide and exchange momentum: dry air and water vapor molecules
outside the volume collide with both dry air and water vapor molecules within it. The difference in the rate of these
collisions across the volume is what determines the vertical pressure gradient $\partial p/\partial z$.
Figure~\ref{fig1} illustrates this basic point.

Since external forces in Eqs.~(\ref{ev}) and (\ref{ed}) are incorrectly specified by \citet{jaramillo18},
Eqs.~(\ref{ev}) and (\ref{ed}) are also incorrect as equations of motion, i.e.
the sum of the forces in the right-hand side of these equations, taken per unit mass, is not equal to accelerations $a_{zv}$ and $a_{zd}$.
Therefore, $F_{vd}$ cannot be retrieved from the condition $a_{zd} =a_{zv} = 0$ and remains unspecified. The statement of
\citet{jaramillo18} summarized by Eq.~(\ref{fvd}) -- that the evaporative does
force $f_e$ introduced by \citet{hess07}
is related to the internal interactions between dry air and water vapor within the considered unit volume of air --
remains unsupported. And with $F_{vd}$ unspecified, the main conclusion of
\citet{jaramillo18} that CIAD {\it ``includes $F_{vd}$ in the vertical motion equation while omitting $F_{dv}$"}
does not have any grounds.

Furthermore, \citet{jaramillo18} provide no evidence that CIAD modifies the equation of vertical motion {\it in any way} (NB: it does not). \citet{jaramillo18} did not quote any equation from our publications that would support their statements.
Rather, they incorrectly attributed their Eq.~(11) to \citet{jetp12} (again without giving an equation number)\footnote{We note that if the last equality in  Eq.~(\ref{vm}) ($= \rho a_z$) is absent, this equation ceases to be the equation of vertical motion. It becomes a definition
of a certain vertical force $F_z$ with an unspecified relation to vertical acceleration. For example, gravity $-\rho g$ is {\it a vertical force}, but
it is not {\it the total vertical force} $F_z$ that determines the vertical air acceleration in (\ref{vm}), i.e. $-\rho g \ne \rho a_z$. We suspect that the vertical force $f_e/\rho$ has been misunderstood by \citet{jaramillo18} as equivalent to the vertical acceleration $a_z$ in the equation of motion.}.

Everywhere in our works the equation of vertical motion is used in the form, using notations of \citet{jaramillo18}, $\rho a_z = -\partial p/\partial z - \rho g$, see, for example, Eq.~(15) of \citet{hess07}, where $\rho a_z = -\partial p/\partial z - \rho g = f_e$, and Eq.~(19) of \citet{acp13},
where $\rho a_z = -\partial p/\partial z - \rho g = 0$ (hydrostatic equilibrium).

\section{CIAD and potential energy}

One key element of CIAD correctly identified by \citet{jaramillo18} is the non-equilibrium vertical distribution of atmospheric water vapor. Due to the condensation that occurs in the rising air and removes water vapor from the gas phase, the negative partial pressure gradient of saturated water vapor is several times larger than the weight of a corresponding amount of moist air. \citet[][their Eqs.~15, 16]{hess07}
proposed that the corresponding vertical force, termed {\it ``evaporative force"} $f_e$, {\it ``drives the global circulation"}:
\begin{equation}\label{fe}
\begin{gathered}
f_e \equiv -\frac{\partial p_v}{\partial z} - \rho_v g = \frac{p_v}{h_c}\left(\frac{h_v - h_c}{h_v}\right),   \\
h_c \equiv \frac{RT^2}{L\Gamma} \ll h_v \equiv \frac{RT}{M_v g},
\end{gathered}
\end{equation}
where $L$ (J~mol$^{-1}$) is the latent heat of vaporization and $\Gamma \equiv - \partial T/\partial z$.  \citet{hess07} backed up this
proposition with the observation that in diverse circulation patterns, irrespective of their size and geometry (e.g. tornadoes versus Hadley circulation), the pressure difference governing air motion is of the order of $10$~hPa -- coinciding, in the order of magnitude, with the partial pressure of water vapor at the surface $z = 0$
\begin{equation}\label{pv}
\Delta p(z) \equiv \int\limits_z^\infty f_e dz \lesssim p_v(z).
\end{equation}
This remarkable universality of atmospheric motions was noted by \citet[][p.~4]{holton04}, but it has not been
reflected in theories of atmospheric circulation.

While \citet{hess07} proposed $f_e$ as a dominant driver of circulation, providing
a testable quantitative framework was left to
subsequent publications. The magnitude of $\Delta p$ (\ref{pv}) (J~m$^{-3}$) was interpreted
as the local store of potential energy available for conversion to kinetic energy \citep[][]{pla09a,pla09b,ijw10}.
A possible analogy is a spring compressed from an equilibrium state with length $h_v$ to
$h_c < h_v$; this spring decompresses in the upward direction
until Hooke's force associated with its deformation (``$-\partial p_v/\partial z$") becomes
balanced by spring's weight (``$-\rho_v g$").

The magnitude of potential energy depends on how the state with minimum potential energy is defined
\citep{lorenz55}. \citet{hess07} considered as such a static atmosphere
where every $i$-th gas with partial pressure $p_i$ and
molar mass $M_i$ has its own scale height $h_i \equiv -p_i/(\partial p_i/\partial z) = RT/M_i g$. However,
in the real atmosphere already in the presence of small vertical motions
(but in the absence of condensation) the air is well mixed in the vertical: its
molar mass $M$ is independent of altitude and all gases have same scale height $h_i = h = RT/Mg$.
Accordingly, in later CIAD publications the definition of the evaporative force (also termed the {\it ``evaporative-condensational"} or
{\it ``condensational"} force) was modified, with $h_v$ in (\ref{fe}) replaced by $h$ \citep[][Eq.~(15)]{jetp12}:
\begin{equation}\label{fc}
f_e \equiv \frac{p_v}{h_c} - \frac{p_v}{h} = -p \frac{\partial \gamma}{\partial z},
\end{equation}
where $\gamma \equiv p_v/p$. This distinct formulation, noted by \citet{jaramillo18}, presumes
that the minimum of condensation-related potential energy characterizes air well mixed in the vertical.
After condensation disturbs the pressure distribution of moist air by removing water vapor from the gas phase,
the air as a whole tends to relax to hydrostatic distribution with the scale height $h$, $p/h = \rho g$.
By analogy, the state with minimum available potential energy as defined by \citet{lorenz55} is
not a static isothermal atmosphere, but an atmosphere with an adiabatic vertical lapse rate, which requires
some motion to be maintained. Defining $f_e$ as in (\ref{fc}) likewise presumes
that some small motion (not generated by condensation) is required to counteract molecular diffusion and maintain a vertical air distribution with  $M = const$ and $h_i = h$.

The key statement of CIAD is that condensation provides power to atmospheric circulation: the rate
at which the kinetic energy of wind is generated is equal to
the rate at which the condensation-related potential energy is released.
The latter rate is equal to the work per unit time $\mathbf{v} \cdot \mathbf{f}_e = w f_e$ of the evaporative force,
where $\mathbf{v}$ and $\mathbf{w}$ are the total and vertical air velocities. It is in this sense that the evaporative force drives winds.
Accordingly, the key equation of CIAD is the equality between
$w f_e$ and the local rate of generation (and, in the steady state, dissipation) of kinetic energy.
For a hydrostatic atmosphere this equation takes the form
\begin{equation}\label{key}
w f_e = -\mathbf{u} \cdot \nabla p,
\end{equation}
where $\mathbf{u}$ is the horizontal velocity ($\mathbf{v} = \mathbf{w} + \mathbf{u}$), see Eq.~(8),
Eq.~(4), Eq.~(17) and Eq.~(5)  of, respectively, \citet{pla09a,pla09b,ijw10,pla11},
Eq.~(16) of \citet{jetp12}, Eq.~(37) of \citet{acp13}. Repeatedly emphasized as
the {\it ``key relationship"}, {\it ``main dynamic equation"} of CIAD etc. \citep[see, e.g.,][]{pla09b,jetp12},
Eq.~(\ref{key}) has escaped notice of \citet{jaramillo18}.

\section{CIAD and dry air}

\citet{jaramillo18} expressed concerns about our treatment of dry air by noting
that CIAD, {\it ``in general, does not address the role of dry air correctly in the mixture, in particular
during the process of condensation. In the real world, the vertical expansion of the water vapor
column due to the difference between its actual and aerostatic scale heights is frustrated by the
dry atmosphere."}

While \citet{jaramillo18} do not provide a quantitative definition of the ``frustration",
one can guess that they refer to the phenomenon considered in detail by \citet{jetp12}
and \citet{acp13}. \citet[][p.~1047]{acp13} wrote: {\it ``when water vapor condenses and its distribution
is compressed several-fold compared to the hydrostatic distribution,
the dry air must be ``{\it stretched}" compared to its
hydrostatic distribution. Only in this case, when the non-equilibrium
deficit of vapor in the upper atmosphere is compensated
by the non-equilibrium excess of dry air, the moist
air as a whole will remain in equilibrium"}. See also on p.~1053:
{\it ``Condensation causes the distribution of vapor $N_v$ to deviate
from the equilibrium distribution. The condition that
moist air as a whole nevertheless remains in equilibrium
causes dry air $N_d$ to also deviate from the equilibrium - but
in the opposite direction to the vapor"}. \citet{jetp12} emphasized that {\it ``condensation changes the vertical distribution of both
water vapor and dry air components"} and described how (see their Eq.~17).

\citet{jaramillo18} neglected these arguments\footnote{A. M. Makarieva brought these arguments to the attention of \citet{jaramillo18}
in a personal email communication to D. Raymond, O. Mesa and A. Jaramillo of 01 November 2017.}
when they state that CIAD is incorrect {\it ``by including $F_{vd}$ in the vertical motion equation
while omitting $F_{dv}$ and not addressing the role of dry air in the mixture. Since these two forces
cancel, their net effect on the moist atmosphere as a whole is zero. Thus, they have no effect
on geophysical fluid dynamics."} As we already discussed above, the argument about $F_{vd}$ is
invalid and thus irrelevant.

Regarding the role of dry air, using the spring analogy, as the spring decompresses upwards it lifts moist air (which is mostly dry air)
to fill the void caused by condensation and to re-establish the hydrostatic equilibrium.
Unlike for water vapor, which is replenished by evaporation, there is no source of dry air at the surface. Hence, as soon as
the dry air has moved upward, the air pressure at the surface drops. This results in a horizontal pressure gradient
to drive horizontal winds with a power $-\mathbf{u} \cdot \nabla p$ equal to the power
$w f_e$ (\ref{key}) of potential energy release during condensation -- the process that has led to the
formation of this pressure gradient. \citet[][p. 1047]{acp13} noted that {\it ``the horizontal pressure gradient produced by condensation
is a direct consequence of hydrostatic adjustment".}

The effects of these processes on atmospheric circulation were illustrated by showing
that Eq.~(\ref{key}) provides a satisfactory quantitative explanation
for the observed wind and pressure profiles in
hurricanes and tornadoes \citep{pla09b,pla11,tor11}. Furthermore, the global integral of Eq.~(\ref{key})
produces an estimate of condensation-driven global circulation that likewise is in a satisfactory
agreement with observations \citep{acp13,jas13,arxiv17}. Since Eq.~(\ref{key}) presumes that condensation is associated with the vertical temperature drop and vertical air motion, a generalization to this
equation was obtained accounting for horizontal temperature gradients \citep{ijw10,pla14}.

\section{CIAD, heat engines and buoyancy}

\citet{jaramillo18} assert some general claims requiring rebuttal.
For example, they state that {\it ``MGH [presumably intended as our team more generally?] never made any serious efforts in presenting the existing theory (or theories) on the maintenance of flows in a moist atmosphere. This is clear from MGH's
difficulty in understanding the production of work in thermally direct circulation, by asserting that
the ascending and descending components of work done by the buoyancy force cancel, resulting
in a significant decrease in the energy released by atmospheric circulations. We showed using the
Kelvin circulation theorem that the latent and sensible heating injected into the circulation at low
levels by surface fluxes have a net positive effect on the circulation."} This is incorrect and misleading.

First, we note that our group has been quite attentive to the {\it ``existing theories
on the maintenance of flows in a moist atmosphere"}. Among our recent works, \citet{jcli15} discussed how
the concept of surface pressure gradients driven by surface temperature gradients
is considerably less robust than commonly thought. \citet{tellus17} presented {\it ``an analytical approach that relates
kinetic energy generation in circulation cells, viewed as heat engines or heat pumps, to surface pressure and temperature
gradients"}. \citet[][]{ar17} discussed how condensation-induced hurricanes relate to the concept of
maximum potential intensity of hurricanes of \citet{em86}.

Second, \citet{jaramillo18} incorrectly attribute the statement that
{\it ``the ascending and descending components of work done by the buoyancy force cancel"} to ``MGH".
This pattern was explained by \citet{goody03}, who convincingly showed that
the {\it ``net positive effect on the circulation"} of surface fluxes can be arbitrarily small.
\citet[][Fig.~2]{goody03} demonstrated that when the radiative cooling of the descending air is small,
the work done by moist convection can be negative. \citet{jaramillo18} ignored these arguments\footnote{The paper of
\citet{goody03} was not quoted by \citet{jaramillo18} despite being brought to their attention by A. M. Makarieva in a personal email communication to A. Jaramillo and O. Mesa of 11 July 2017.}
and, based on the work of \citet{pa11}, considered a specific example where the radiative
cooling in the upper atmosphere is such that the work of the cycle is positive.
However, the lower limit for the efficiency of the buoyancy-driven circulation cannot be, and has not been, obtained
from the heat engine considerations. Rather than {\it ``readily explaining"} circulation over forests,
\citet{jaramillo18} did not present any evidence that their examples are quantitatively relevant to the real atmosphere\footnote{Note also that
Eq.~(10) of \citet{jaramillo18} represents a formal replacement of variables: in $F_z \equiv -\partial p/\partial z - \rho g$
pressure $p$ is replaced by a combination of $\rho_d$, saturated mixing ratio $r^*_v$ and temperature $T$ using the ideal gas law
and the equation of moist adiabat. The resulting expression for $F_z$ does not contain any information regarding the role of condensation in atmospheric dynamics since  all these new variables, as $p$ in the original expression, remain unspecified.
Likewise, the definitions of the evaporative force (\ref{fe}), (\ref{fc}) by themselves do not contain any information about
condensation being a circulation driver. It is only Eq.~(\ref{key}) that does.}. In contrast, applying CIAD
specifically to the Amazon region, \citet{jhm14} demonstrated that the theoretical predictions of CIAD
agree with observations of surface pressure gradients and velocities.

Finally, \citet{jaramillo18} juxtapose CIAD to the consideration of the atmosphere as {\it ``a heat engine
that produces mechanical work by transporting energy from warm to cold regions"}. It is a false opposition.
Since atmospheric air circulates between the warm surface and the cold upper atmosphere,
it is always possible to describe this motion as a thermodynamic cycle, irrespective
of what drives it: condensation, buoyancy or something else.

The dynamic and thermodynamic approaches to atmospheric circulation
combine in the relationship between pressure gradient and heat input.
Indeed, in a steady-state, to receive heat from the ocean surface air must move (otherwise
its temperature would rise) and this requires a pressure gradient.
Without specifying this pressure gradient, it is not possible to
determine the amount of heat received and, hence, the work performed within the thermodynamic cycle.

For example, in the thermodynamic cycles considered by \citet{pa11}
(whose ideas \citet{jaramillo18} reproduce in considerable detail) -- the ``steam cycle"
(only latent heat consumed from the ocean) and the ``mixed cycle"  (both latent and sensible heat consumed) --
the surface pressure difference is an {\it external parameter}. This surface pressure difference, which drives
surface winds, cannot be retrieved from the heat engine approach. In consequence, \citet{pa11} could not uniquely relate work
output to the moisture supply (condensation rate) within the cycle: the relationship relied on the unknown or {\it a priori}
postulated surface pressure difference. (Specifically, \citet{pa11} set the surface pressure difference to zero for his
``steam cycle" and left it unspecified (determining the sensible heat flux) for his ``mixed cycle").
Relating work output to moisture input requires an extra equation -- a constraint governing the dynamics of the boundary layer.
Equation~(\ref{key}) of CIAD provides such a constraint.

The mere existence of a heat source and a heat sink does not guarantee that work will be performed.
There must be a dynamic system which can convert heat into potential energy and work
\citep[see discussion by][]{dhe10}.
E.g. a spring attached to a piston in a cylinder accumulates potential energy while the gas within the cylinder expands and then pushes the piston
back to compress the gas.
In this context, the statements of \citet{pa11,pa15}
and colleagues that the water cycle {\it limits} [i.e. reduces] {\it the work output of the atmospheric heat engine} [because of
irreversible processes associated with phase transitions] can be compared to the statement
that since the spring is not ideal and has internal friction, the spring reduces the {\it maximum} possible output of useful work within the cycle.
This is correct, but one should not read this statement as {\it without the spring (or the water cycle)
the work would be larger than it is in its presence}. Rather, it is crucial to note that without a spring there would be no work at all.
While the standard theory identifies the pressure gradients caused by differential heating
as such a spring, we propose instead that the main dynamic mechanism providing power
to the Earth's circulation is the non-equilibrium vertical distribution of water vapor caused by condensation.

The equations of hydrodynamics allow for a solution where despite the same horizontal differential heating the power
of atmospheric circulation on Earth is zero (the atmosphere is in geostrophic balance and no kinetic energy is generated).
In current atmospheric models a non-zero power is achieved by fitting the parameters of turbulent friction.
CIAD by constraining atmospheric power output, Eq.~(\ref{key}), actually guides the parameterization of turbulence
(which in current models is not explicitly related to condensation).

It is from our observation that atmospheric power is well predicted by CIAD that we conclude that the net effect of
temperature gradients (differential heating) is small. This implies that on a dry Earth the power of atmospheric circulation would be smaller than now.  More importantly, removing major sources of water vapor, e.g. through large-scale deforestation, will influence atmospheric circulation, modify ocean-to-land moisture transport and impact the terrestrial water cycle \citep{hess07,mgl13,nobre2014,sheil2018}.
Independent observation-based studies testify in favor of a significant impact of vegetation cover on ocean-to-land circulation and moisture import  \citep[e.g.,][]{Levermann2009,ch10,andrich13,poveda14,Herzschuh2014,Levermann2016,Boers2017}.
These scattered studies currently lack a unifying theoretical framework.
One reason is that current circulation models do not appear to support abrupt changes in air circulation
following changes in the functioning of vegetation cover \citep[e.g.,][]{Boos2016b}.
However, if modeled turbulence could be re-parameterized so as to account for CIAD,
we expect the simulated atmospheric reactions to vegetation removal to be more realistic.
We thus welcome discussion and strongly advocate the rigorous and
focused (re)appraisal of the implications of forest and land cover change for atmospheric circulation and moisture transport.

%
\acknowledgements

This work is par\-ti\-al\-ly supported by  the University of California Agricultural Experiment Station,
Australian Research Council project DP160102107 and the CNPq/CT-Hidro - GeoClima project Grant~404158/2013-7.

\bibliographystyle{copernicus}

\end{document}